\documentclass[%
 aip,
 jmp,%
 amsmath,amssymb,
 reprint,%
]{revtex4-1}

\usepackage{graphicx}
\usepackage{dcolumn}
\usepackage{bm}

\begin{document}


\title{A laser heterodyne polarimeter for birefringence measurement}

\author{Harold Hollis}
\author{Gabriel Alberts}
\author{D. B. Tanner}
\author{Guido Mueller}
\affiliation{Department of Physics, University of Florida, PO Box 118440, Gainesville, Florida 32611, USA}

\date{\today}

\begin{abstract}
We introduce a laser heterodyne polarimeter designed for the precision measurement of sub $\mu\mathrm{rad}$ differential phase shifts due to birefringence. The polarimeter will be used in an initial testbed for a potential future vacuum magnetic birefringence experiment at DESY. This experiment would use the 212-m-long ALPS magnet string. The vacuum magnetic birefringence signal will be amplified inside a high finesse optical cavity before it is sensed.
This paper describes the polarimeter, initial results, and systematic error sources which still have to be minimized before the vacuum birefringence experiment can be realized.
\end{abstract}
\maketitle
\section{Introduction}
Vacuum magnetic birefringence (VMB), an effect predicted by Heisenberg and Euler in the 1930s, is a nonlinear QED phenomenon in which the index of refraction in vacuum is a function of the direction and amplitude of a static magnetic field. VMB manifests itself in the accumulation of a phase difference between light polarized parallel and perpendicular to the magnetic field on account of refractive index differences for light polarized parallel ($n_\parallel$) and perpendicular ($n_\perp$) to the field
\begin{equation}
    \gamma = 2\pi \left(n_{\parallel}-n_{\perp}\right) \frac{L}{\lambda}
\end{equation}
where $L$ is the propagation distance and $\lambda$ is the wavelength of the light. The refractive index difference is extremely small ($\Delta n = 4\times 10^{-24} \mbox{ T}^{-2}\times B^2$); obtaining measurable phase change requires a very strong magnetic field, a long travel distance, and the employment of high finesse optical cavities. 

Attempts to measure VMB continue to improve in sensitivity but are plagued by systematic errors and thermal noise often associated with the mirrors and their coatings which form the optical cavities  \cite{DellaValle2016, Zavattini2018, Hartman2019}.

The upcoming ALPS II experiment\cite{Spector2019} at DESY uses two 106 m long strings, each made using 12 HERA magnets ($5.3\ \mbox{T}$), to search for axion-like particles. This setup provides the largest $B^2\cdot L$ in the world. Similar to \cite{Cadene2014} and \cite{DellaValle2013}, a high finesse optical cavity amplifies the path length potentially by five orders of magnitude. However the much longer distance between the mirrors promises to reduce the relative impact of systematic effects and thermal noise from the mirrors and their coatings.

The differential phase change (DPS) in such an experiment is
\begin{equation}
    \Gamma \propto \left[\frac{1\ \mu\mbox{m}}{\lambda}\right] \left[\frac{L}{200\ \mbox{m}}\right]
    \left[\frac{B^2}{25\ \mbox{T}^2}\right]\left[\frac{\mathcal{F}}{1.25\times 10^5}\right]\cdot 10\ \mbox{nrad} 
\end{equation}
where $\mathcal{F}$ is the finesse of the cavity. Once fully commissioned, our laser heterodyne polarimeter (LHP) is expected to have the capability to measure differential phase changes at this level.

Our technique is based on an idea initially published by Hall\cite{Hall2000} where it was proposed to measure cavity mirror birefringence at the sub-ppm level with frequency metrology. The mirror is probed with two orthogonally polarized laser fields for which the polarization axes are rotated with constant angular speed. The reflected fields then contain an oscillating phase relationship with an amplitude equal to the DPS of the birefringent mirror. In its current version, the LHP has succeeded in measuring differential phase changes at the sub 100 nrad level. 
\begin{figure}[htbp] 
\centering
\includegraphics[width=\linewidth]{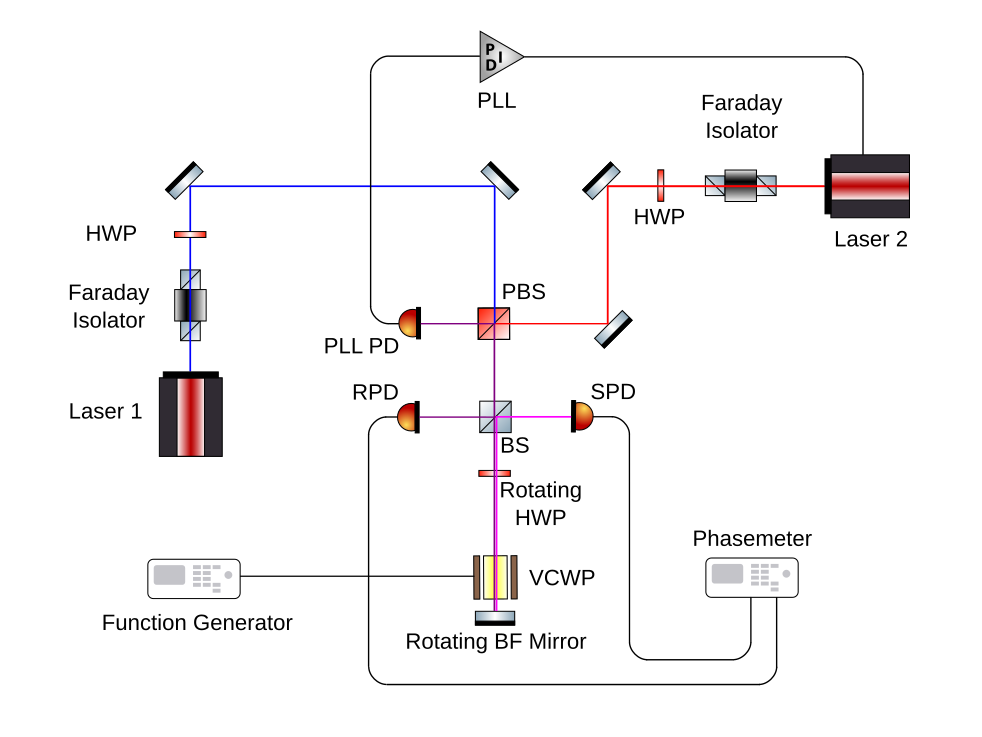}
\caption{Schematic diagram of the LHP (Not shown: optics for setting the light field amplitudes seen by the photodiodes and polarizing optics before each photodiode for aligning field directions). HWP and VCWP are half-wave plate and voltage controlled wave-plate, BS and PBS are power and polarizing beamsplitter, and PLL PD, RPD, and SPD are phase-lock loop, reference and signal photodiode respectively.}
\label{fig:schematic}
\end{figure}
\section{Operational Overview}
\label{sec:overview}
A schematic of the experimental apparatus is shown in Fig.\ref{fig:schematic}. The LHP uses two single frequency lasers \cite{Kane:85} that are phase locked with a phase-locked loop (PLL) to a constant angular frequency offset $\Omega$; the offset frequency is chosen to be within the bandwidth of the photodetectors. A Faraday isolator at the output of each laser prevents backscattered light from entering the laser while a HWP which follows each Faraday isolator is set to balance the power in the $s$ and $p$ polarizations.

The two laser fields are combined with a polarizing beam splitter (PBS). The PBS directs the $s$ polarized field from laser 1 $(s_1)$ and the $p$ polarized field from laser 2 $(p_2)$ towards the PLL PD to phase lock the lasers. A HWP, set to $22.5^\circ$, rotates both fields by $45^\circ$, and a PBS selects the $s$-polarized component from each laser field to be sent to the PLL PD. 

The $s_2$ and $p_1$ fields are directed toward a power beam splitter (BS). The reflected beam from the BS is the \textit{reference beam} while the transmitted beam is the \textit{probe beam}. The reference beat signal is the reference input to a phasemeter \cite{Moku:Lab} (PM).

The probe beam propagates through a HWP rotating with angular speed $\omega_\pi$, through a VCWP, and then reflected (cyan line) from the birefringent mirror, which can be rotated with angular speed $\omega_m$, to propagate back through the VCWP, rotating HWP, and into the BS. The reflected probe beam from the BS produces a \textit{phase modulated} beat signal at the signal photodetector (SPD). This probe beat signal is the probe input to the PM. The PM measures the relative phase of the reference and probe inputs and produces a phase difference time series which is recorded to a file. The recorded PM data is analyzed to extract the mirror and VCWP BF signals.

As is shown in the appendix, constant VCWP BF and mirror BF will result in phase modulation of the probe beat signal at angular frequencies $4\,\omega_\pi$ and $4\,\omega_\pi - 2\,\omega_m$ respectively. Driving the VCWP with a sinusoidal voltage at angular frequency $\omega_{\nu}$ will produce sideband components at angular frequencies $4\,\omega_\pi\pm \omega_{\nu}$. It is also shown that any static departure $\epsilon$ from ideal half-wavelength retardation in the rotating HWP will produce a component at $2\,\omega_\pi$, and that a time dependent $\epsilon(t)$ due to non-zero HWP angle of incidence (AOI) will produce a component at $4\,\omega_\pi$.
\section{Experimental Setup and Instrumentation}
For our LHP implementation, we use $1064\,\mathrm{nm}$ lasers locked with a PLL to a frequency offset of $5\,\mathrm{MHz}$. A Thorlabs DDR05 motorized rotation stage is used to rotate a HWP at a rate of $1800^{\circ}$ per second $(5\,\mathrm{Hz})$. A Zaber RSB060AD motorized rotation stage is used to rotate the mirror-under-test (MUT) at $f_m \leq 8.33\,\mathrm{Hz}$ in the opposite sense of the HWP rotation. For these rotation rates, we expect to see the 1st order HWP imperfection signal at $10\,\mathrm{Hz}$ and the mirror BF signal at $20\,\mathrm{Hz} + 2\,f_m$. The VCWP, a Newport 4104NF broadband EOAM, is driven by a function generator (FG) through a $62\,\mathrm{dB}$ attenuator.

For data acquisition, we use the Liquid Instruments Moku:Lab \cite{Moku:Lab} running the phasemeter instrument. The phasemeter samples the phase of the two $5\,\mathrm{MHz}$ input signals at a rate of $1\,\mathrm{MHz}$ and the resulting phase estimate time series are decimated to a rate of $122.07\,\mathrm{Hz}$ for saving to disk. The recorded data are processed with Matlab.

To extract the phase modulation signals due to the VCWP BF, mirror BF, and HWP imperfection, and to reduce common mode phase noise, the difference of the recorded phase estimate time series for each channel is used for the data analysis. The precise frequency and amplitude of the expected signals is then determined using a combination of frequency and time domain techniques.

During our testing, a compound zero order HWP was used for the rotating HWP.  We consistently measured a $10\,\mathrm{Hz}$ signal amplitude of $\approx 15\,\mathrm{mrad}$ indicating a static retardation error of $\approx 7.5\,\mathrm{mrad} = \frac{\lambda}{838}$. The retardation error's quadratic dependence on AOI was confirmed and measured to be approximately $\xi(\alpha) \approx 10\,\alpha^2\,\mathrm{mrad}/\mbox{deg}^2$ where $\alpha$ is in degrees.

For our initial tests, the MUT was an off-the-shelf broadband dielectric mirror. This mirror's BF consistently produced a signal amplitude of approximately $1.5\,\mathrm{mrad}$.  We did note that removing and replacing this mirror in the rotation stage would change the measured DPS slightly which might be due to some stress induced BF.

For a wavelength of $1064\,\mathrm{nm}$, the VCWP produces $\pm\lambda/2$ retardation for an input voltage of $\pm 320\,\mathrm{V}$. With the FG set to its minimum amplitude $(5\,\mathrm{mV})$, the minimum VWCP drive amplitude is $4.0\,\mu\mathrm{V}$ producing a sinusoidal VCWP DPS of amplitude $39\,\mathrm{nrad}$. 

\section{Data Analysis}
For the detection of low-level sinusoids in noise, we utilize a technique described in \cite{Bush2019} as $Z$ demodulation. The phasemeter time series is I/Q demodulated at the expected signal frequency, the cumulative sum of the I and Q channels are squared and summed and then normalized by the squared length of the time series. The resulting series, $Z(\tau)$, is essentially the magnitude squared of the (normalized) DFT (evaluated at the demodulation frequency) as function of the integration time $\tau$.

\begin{figure}[!htbp]
\includegraphics[width=\linewidth]{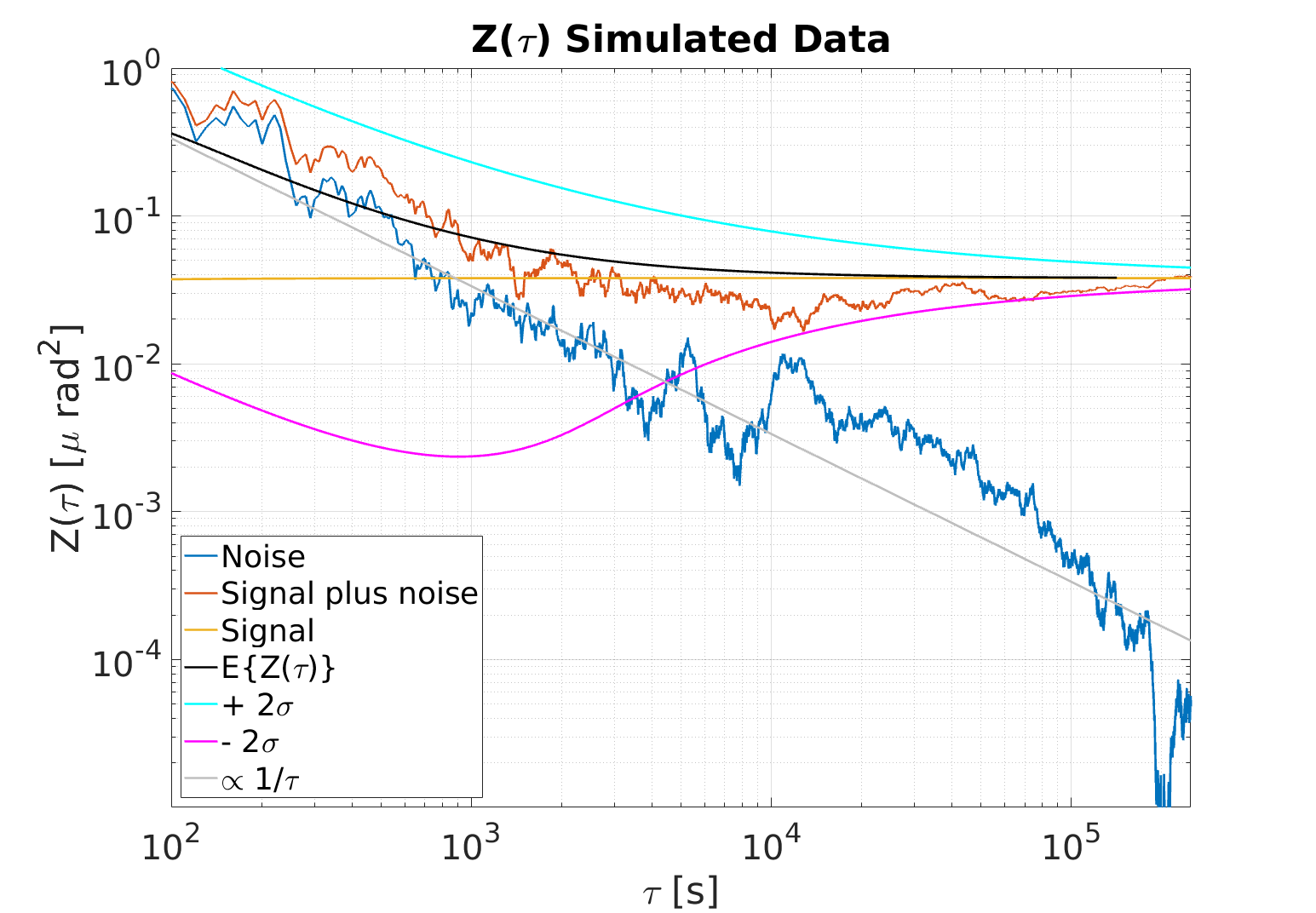}
\caption{$Z(\tau)$ for simulated data. For noise (blue), the expectation of $Z(\tau)$ follows a trend line (gray) proportional to $1/\tau$ while a pure signal (green) of amplitude $A$ gives a flat line at $A^2 / 4$. For signal plus noise (red), $Z(\tau)$ initially follows the noise trend and then approaches the pure signal line. The expectation (black) and $\pm 2\,\sigma$ bounds (cyan, magenta) of $Z(\tau)$ for signal plus noise are also plotted.}
\label{fig:Z(tau)}
\end{figure}
 
For a pure sinusoidal signal of amplitude $A$, $Z(\tau)$ goes to $A^2 / 4$. For random noise, $Z(\tau)$ is a random walk around a trend line proportional to $1/\tau$. In Fig. \ref{fig:Z(tau)}, $Z(\tau)$ is plotted for a simulated signal, simulated noise, and for the sum of the signal and noise. We also plot the expectation value and $\pm 2\,\sigma$ bounds of $Z(\tau)$ for signal plus noise.

\section{Results for VCWP BF}
\begin{figure}[htbp]
\centering
\includegraphics[width=\linewidth]{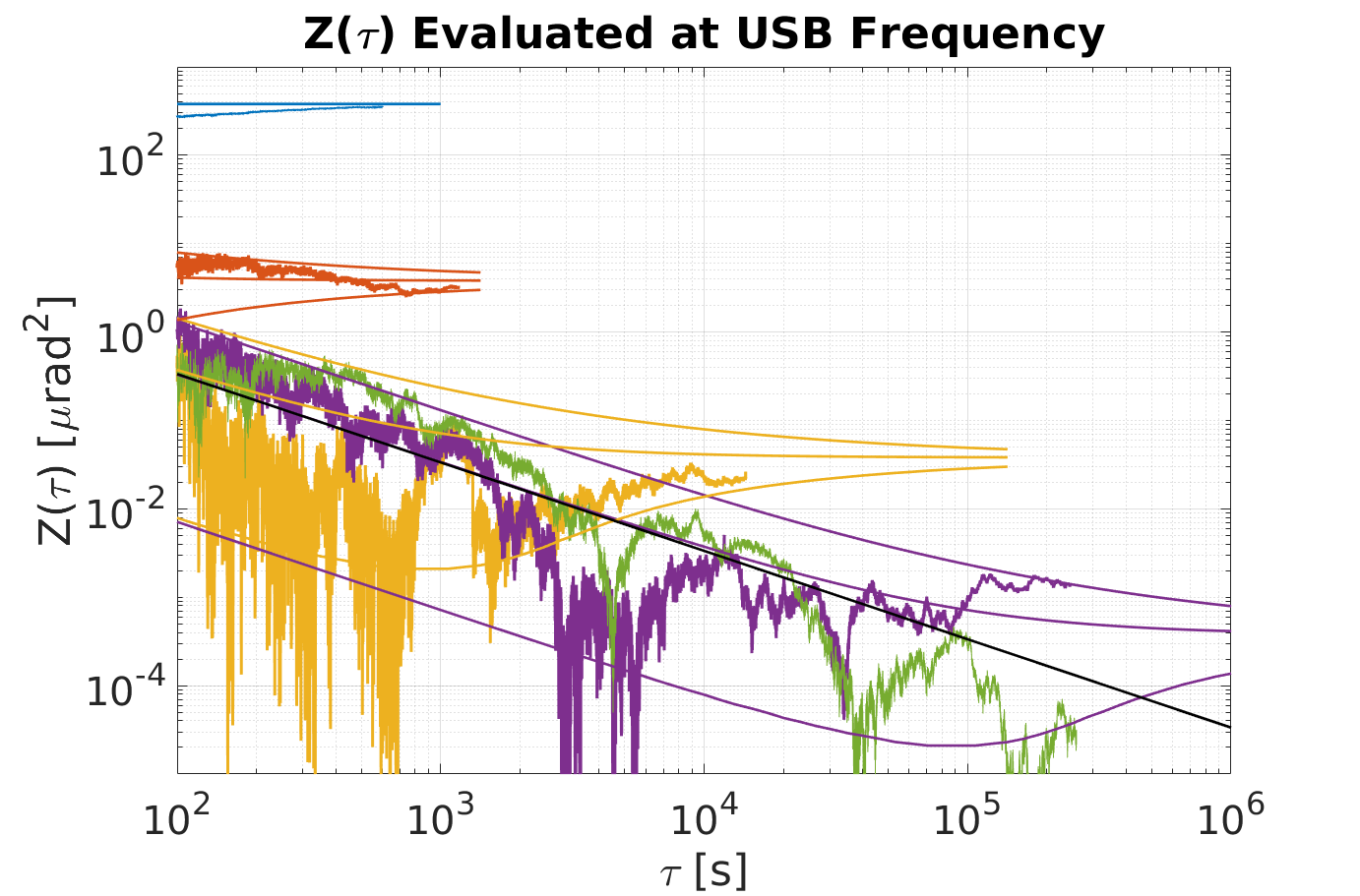}
\caption{$Z(\tau)$ at USB frequency for four VCWP drive levels.  From top to bottom, the expected VCWP DPS amplitudes are: $39\,\mu\mathrm{rad},\,3.9\,\mu\mathrm{rad},\,390\,\mathrm{nrad},\,39\,\mathrm{nrad}$ and zero (FG disconnected). Expectation and $\pm 2\,\sigma$ bounds are plotted along with the $Z(\tau)$ for each data run with drive. The black line is the trend line for noise only.}
\label{fig:Z(tau)_USB}
\end{figure}

\begin{figure}[htbp]
\centering
\includegraphics[width=\linewidth]{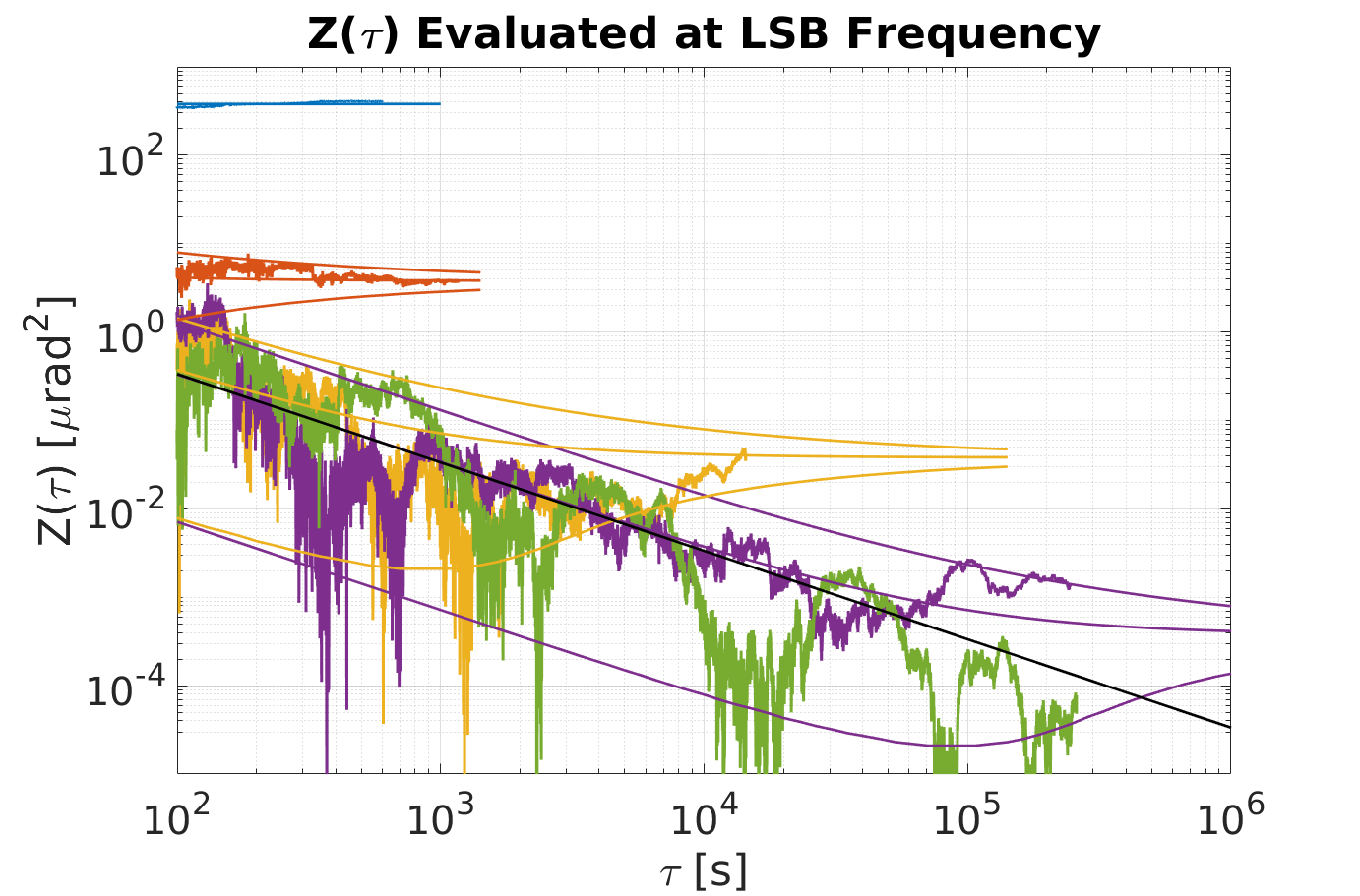}
\caption{$Z(\tau)$ at LSB frequency}
\label{fig:Z(tau)_LSB}
\end{figure}

For the VCWP BF tests, the mirror was not rotated and the FG frequency was set to $f_\nu \approx 6.67\,\mathrm{Hz}$.  Data was taken at FG amplitude settings corresponding to VCWP DPS amplitudes of $39\,\mu\mathrm{rad},\,3.9\,\mu\mathrm{rad},\,390\,\mathrm{nrad},\,39\,\mathrm{nrad}$ as well as with the FG disconnected.  The results at the USB and LSB frequencies are plotted in Fig. \ref{fig:Z(tau)_USB} and Fig. \ref{fig:Z(tau)_LSB} respectively.

The expectation of $Z(\tau)$ along with the $\pm 2\,\sigma$ bounds are plotted along with $Z(\tau)$ for each data run with drive. The approach of $Z(\tau)$ to the expectation is apparent for the $390\,\mathrm{nrad}$ and larger VCWP DPS amplitudes. While it is clear that there is a signal present for the $39\,\mathrm{nrad}$ amplitude run, the $\approx 68\,\mathrm{hr}$ integration time for this run is too short to verify the approach of $Z(\tau)$ to the expectation. The $Z(\tau)$ for the data run without drive confirms that there is no significant spurious signal at the USB and LSB frequencies.     

\section{Results for Rotating Mirror BF}

\begin{figure}[htbp]
\centering
\includegraphics[width=\linewidth]{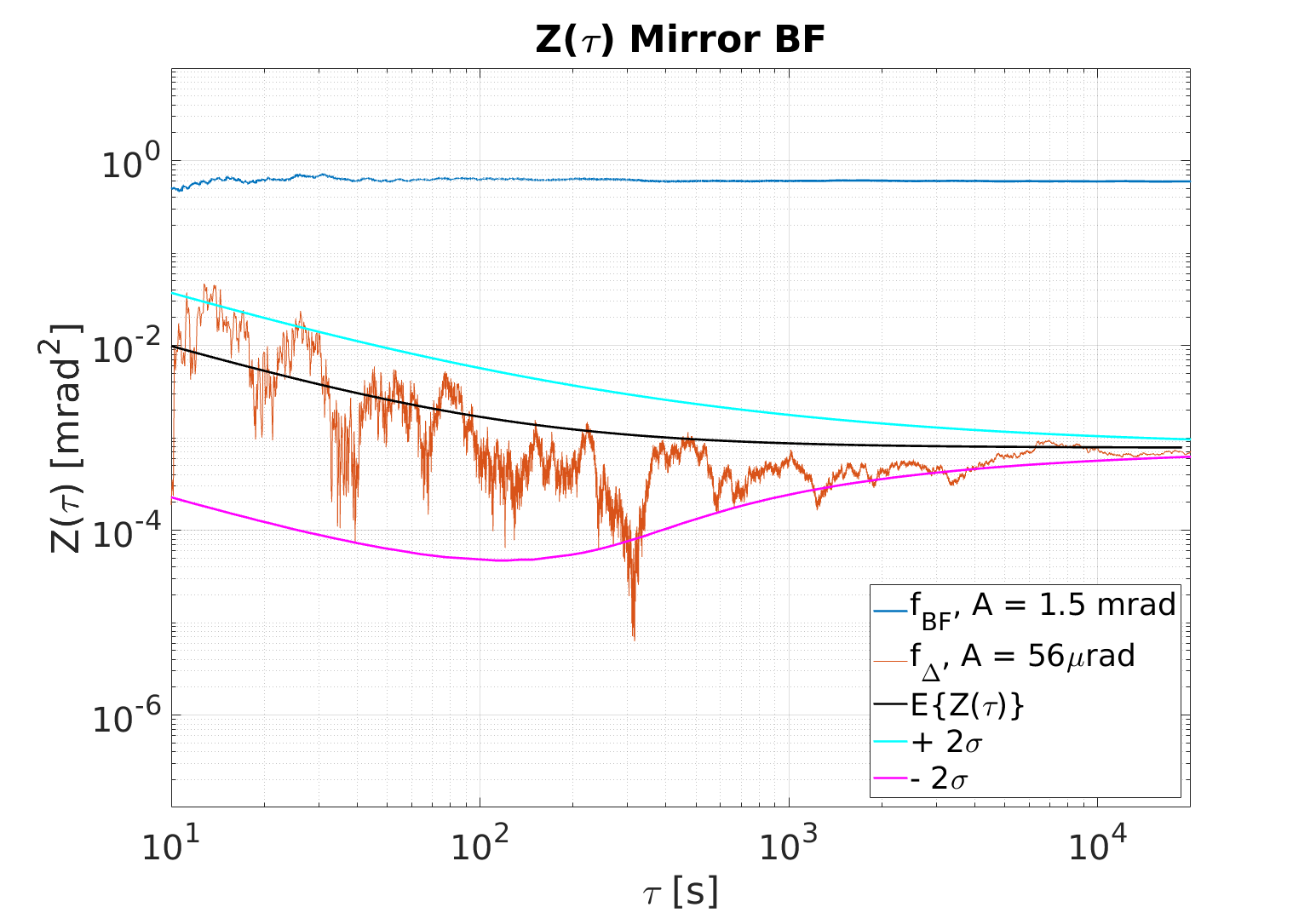}
\caption{$Z(\tau)$ evaluated at $f_{MBF} \approx 26.25\,\mathrm{Hz}$ and at the difference frequency $f_\Delta \approx 13.75\,\mathrm{Hz}$. The presence of a $56\,\mu\mathrm{rad}$ signal at the difference frequency implies a spurious signal source that is still under investigation.}
\label{fig:Z_MBF}
\end{figure}

For the mirror BF tests, we removed the VCWP and rotated the MUT with rotational frequency $f_m \approx 3.125\,\mathrm{Hz}$ in the opposite sense of the HWP rotation. Since non-zero HWP AOI produces a signal at the same frequency as mirror BF when the MUT is not rotating, rotating the MUT moves the mirror BF signal frequency to $f_{MBF} \approx 26.25\,\mathrm{Hz}$. In Fig. \ref{fig:Z_MBF}, $Z(\tau)$ evaluated at $f_{MBF}$ is plotted.  Also plotted is $Z(\tau)$ at the difference frequency $f_\Delta \approx 13.75\,\mathrm{Hz}$. Ideally, there should be no signal at $f_\Delta$. The detection of a $56\,\mu\mathrm{rad}$ signal there implies a spurious signal source.

To investigate this further, we replaced the MUT with a LIGO ETM witness mirror.  We expect the DPS for this mirror to be of the order of $1\,\mu\mathrm{rad}$. For a 53 minute data run, we measured $26.3\,\mu\mathrm{rad}$ and $30.5\,\mu\mathrm{rad}$ at $f_{MBF}$ and $f_\Delta$ respectively which suggests that this mirror's BF signal at $f_{MBF}$ is dominated by a spurious signal there.

\section{Conclusion and Outlook}
With our LHP experiment, we have measured the $1.5\,\mathrm{mrad}$ DPS from a birefringent dielectric mirror as well as sub $\mu\mathrm{rad}$ sinusoidal DPS from a birefringent VCWP. Our measurements confirm the theoretical calculations for the mirror BF signal at $f_{MBF} = 4\,f_\pi - 2\,f_m$, the static HWP error signal at $2\,f_\pi$, the HWP AOI signal at $4\,f_\pi$, and the sinusoidal VCWP BF sidebands at $4\,f_\pi\pm f_\nu$. We continue to investigate the source of the spurious signals at $4\,f_\pi\pm 2\,f_m$ which ultimately limit the sensitivity of the instrument to mirror BF. Simulations suggest that wobble in both the rotating HWP and MUT rotation stages is the primary source of these spurious signals. Sensing and actively controlling this wobble over long periods of time will likely be required to significantly reduce the spurious signal and measure the birefringence of high quality mirrors directly. 

\appendix
\setcounter{secnumdepth}{0}
\section{Appendix - Theoretical Calculations}
In the following, we show that VCWP BF and mirror BF will result in phase modulation of the probe beat signal at angular frequencies $4\,\omega_\pi$ and $4\,\omega_\pi - 2\,\omega_m$ respectively. Departure from ideal half-wavelength retardation in the rotating HWP will produce a component at $2\,\omega_\pi$ while non-zero HWP AOI produces a component at $4\,\omega_\pi$. We assume ideal polarizing and power beam splitters. The Jones vector representation of the combined, orthogonally polarized, phase-locked laser fields is given by:
\begin{equation}
\tilde{\mathbf{E}} = \begin{pmatrix}
1\\
0\end{pmatrix}A_p\,e^{i\omega t} + \begin{pmatrix}
0\\
1\end{pmatrix}A_s\,e^{i(\omega + \Omega)t}
\end{equation}
where the amplitudes $A_p$ and $A_s$ are taken to be real.  The reflected field from the BS is given by 
\begin{equation}
\tilde{\mathbf{E}}_R = \frac{1}{\sqrt{2}}\left( \begin{pmatrix}
1\\
0\end{pmatrix}A_P + \begin{pmatrix}
0\\
1\end{pmatrix}A_s\,e^{i\Omega t} \right)\,e^{i\omega t}
\end{equation}
The AC intensity of the field incident on the RPD, after passing the HWP-PBS combination, is then
\begin{equation}
\tilde{I}_{RPD} = \frac{1}{2}A_pA_s\cos\Omega t
\end{equation}
Following the Siegman convention \cite[p.~406]{Siegman}, the transmitted field from the BS is given by
\begin{equation}
\tilde{\mathbf{E}}_T = \frac{i}{\sqrt{2}}\left( \begin{pmatrix}
1\\
0\end{pmatrix}A_P + \begin{pmatrix}
0\\
1\end{pmatrix}A_s\,e^{i\Omega t} \right)\,e^{i\omega t}
\end{equation}
This field propagates through an ideal HWP rotated at angle $\theta_\pi$, through the VCWP with DPS $\nu$, reflects from the mirror with reflection phases $\phi_x,\,\phi_y$ rotated at angle $\theta_m$, and propagates back through the VCWP and HWP. Denote the mirror DPS as $\delta \equiv \phi_x - \phi_y$. Working to 1st order in $\delta$ and $\nu$, the Jones matrix representation for this combination is (up to an irrelevant phase)
\begin{equation}
\mathbb{1} + i
\begin{pmatrix}
\delta\cos^2\gamma + 2\nu\sin^2 2\,\theta_\pi & \frac{\delta}{2}\sin 2\gamma - \nu\sin 4\,\theta_\pi \\
\frac{\delta}{2}\sin 2\gamma - \nu\sin 4\,\theta_\pi & \delta\sin^2\gamma + 2\nu\cos^2 2\,\theta_\pi
\end{pmatrix}
\end{equation}
where $\gamma \equiv 2\,\theta_\pi - \theta_m$. After reflection at the BS and propagation through the HWP-PBS combination, the AC intensity of the field incident on the SPD is then
\begin{align}
\tilde{I}_{SPD} & = \frac{1}{4}A_pA_s\left[ \cos\Omega\,t + \left( \delta\cos 2\gamma - 2\nu\cos 4\,\theta_\pi \right) \sin\Omega\,t \right]\nonumber\\
&\quad\approx \frac{1}{4}A_pA_s\cos\left(\Omega t - \phi \right)
\label{ISPD}
\end{align}
where $\phi = \delta\cos 2\gamma - 2\nu\cos 4\,\theta_\pi$. Rotating the HWP with constant angular speed $\omega_\pi$ such that $\theta_\pi = \omega_\pi\,t$ will result in phase modulation of the SPD beat signal at an angular frequency of $4\,\omega_\pi$ with an amplitude of $\left(\delta^2 + 4\nu^2 - 4\delta\nu\cos 2\,\theta_M\right)^{1/2}$.  For the $\nu = 0$ case, the phase modulation amplitude is just the mirror DPS:
\begin{equation}
\phi_{\nu=0} = \delta \cos 2\gamma = \delta \cos \left( 4\,\omega_\pi\,t - 2\,\theta_m \right)
\end{equation}
If the VCWP DPS is sinusoidal, $\nu(t) = \nu_0\cos\left( \omega_\nu t\right)$, the VCWP BF signal appears in sidebands of the mirror BF signal frequency:
\begin{equation}
\phi = \delta\cos\left( 4\,\omega_\pi t - 2\,\theta_m \right) - \nu_0\cos\left[ \left( 4\omega_\pi\pm\omega_\nu \right)t \right]
\end{equation}

We model an imperfect rotating HWP by setting the retardation to $\pi + \epsilon$.  Working to first order in $\epsilon$, $\delta$, and $\nu$, an additional term of order $\epsilon$ is added to equation (\ref{ISPD}) above:
\begin{equation}
\frac{\epsilon}{2}A_pA_s\cos 2\,\theta_\pi\sin\Omega t 
\end{equation}
and then the phase becomes
\begin{equation}
\phi = \delta\cos 2\gamma - 2\nu\cos 4\,\theta_\pi - 2\epsilon\cos 2\,\theta_\pi
\end{equation}
As before, for $\theta_\pi = \omega_\pi\,t$ and $\nu = 0$ the time dependent phase term becomes
\begin{equation}
\phi_{\nu=0} = \delta \cos 2\gamma = \delta \cos \left( 4\,\omega_\pi\,t - 2\,\theta_m \right) - 2\epsilon\cos 2\,\omega_\pi\,t
\end{equation}
The rotating imperfect HWP modulates the phase at an angular frequency of $2\,\omega_\pi$ with an amplitude of $2\epsilon$.  It's worth going to second order in $\epsilon$ to find the following additional term
\begin{equation}
-\frac{\epsilon^2}{4}A_pA_s\left(1 + \cos 4\theta_\pi\right)\cos\Omega\,t
\end{equation}
so that the phase becomes
\begin{equation}
\phi_{\nu=0} \approx \delta\cos 2\,\gamma - 2\epsilon\cos 2\,\theta_\pi - \frac{\epsilon^3}{3}\cos 6\,\theta_\pi
\end{equation}
which shows that there is no HWP retardation error signal at the mirror BF signal signal frequency.

A real HWP exhibits a retardation that is quadratically dependent on the angle of incidence $\alpha$.  When the AOI is with respect to the slow (fast) axis, the retardation is maximum (minimum) and so the retardation oscillates through a complete cycle as the HWP rotates through $180^\circ$.  That is, if $\theta_\pi = \omega_\pi\,t$, the \textit{time dependent} retardation error $\epsilon(t)$ oscillates with fundamental frequency $2\,\omega_\pi$. It follows that for a small, non-zero AOI

\begin{equation}
\epsilon(t) = \epsilon_0 + \xi(\alpha)\cos 2\,\omega_\pi\,t
\end{equation}
\begin{equation}
2\epsilon(t)\cos 2\,\theta_\pi = 2\epsilon_0\cos 2\,\omega_\pi\,t + \xi(\alpha)\left[1 + \cos 4\,\omega_\pi\,t\right]
\end{equation}
which shows that there is an HWP AOI signal at $4\omega_\pi$ that limits the sensitivity to a mirror BF signal at this frequency.  The time dependent phase at $4\,\omega_\pi$ is then given by
\begin{equation}
\delta \cos \left( 4\,\omega_\pi\,t - 2\,\theta_m \right) - \xi(\alpha)\cos 4\,\omega_\pi\,t
\end{equation}
This suggests that the mirror BF signal can be moved away from the HWP AOI signal by rotating the mirror, $\theta_m = \omega_m\,t$, so that the mirror BF signal frequency becomes 
\begin{equation}
\omega_{BF} = 4\omega_\pi - 2\omega_m
\end{equation}
\section{Funding}

National Science Foundation (NSF) (1505743); Heising-Simons Foundation (2015-154)
\bibliography{references}
\end{document}